\documentclass[superscriptaddress,preprint]{revtex4}
\usepackage{mathrsfs}
\usepackage{amssymb}
\usepackage[tbtags]{amsmath}
\usepackage{graphicx}
\usepackage{epsfig,graphicx,times}
\usepackage{color}
\usepackage{subfigure}

\begin{document}
\title{Nonclassical effects in geodesic motion}
\author{M. Zhang\footnote{zhangmiao079021@163.com}}
\affiliation{School of Physics, Southwest Jiaotong University,
Chengdu 610031, China}
\author{Y. Wang\footnote{qubit@home.swjtu.edu.cn}}
\affiliation{School of Physics, Southwest Jiaotong University,
Chengdu 610031, China}
\date{\today}
\begin{abstract}
Gravity gradient is known as a serious systematic effect in atomic tests of the universality of free fall, where the initial central position and velocity of atoms need to be exactly controlled. In this paper, we study quantum free fall with high-order gravity gradients. It is shown that the cubic terms in the Newtonian potential shall generate a new phase shift in atom interferometers, which depends on the position and velocity uncertainties of the incident atoms. We further investigate the nonclassicality of free fall and show that, due to the cubic potential, the gravitational Wigner equation in phase space of position and velocity is different from the classical Liouville equation. There exists a mass-dependent correction in the dynamical equation regardless of initial state. Nevertheless, this is just a quantum mechanical effect of microparticles, which does not violate equivalence principle that inertia mass is equal to the gravitational mass.
\end{abstract}

\maketitle
\section{Introduction}
By using some modern techniques, such as neutron interferometer~\cite{COW}, atom interferometer (AI)~\cite{ChuAPB,ChuG,ChuR}, and the optical atomic clocks~\cite{Wineland,Car,clocks-RMP}, gravity near Earth's surface has been measured very precisely.
Using AI, the gravitational acceleration was measured in an accuracy of $10^{-9}g$, with $g\approx 9.8\,\text{m}/\text{s}^2$~\cite{ChuG}. Using optical atomic clocks, the gravitational time dilations due to the height change of less than $1$ meter were observed~\cite{Wineland}. In addition to interference, the bound state is also a typical nonclassical effect predicted by quantum mechanics. Several experiments~\cite{G-bound0,G-bound1,G-bound2} have demonstrated such phenomena in gravitational potential via the reflecting neutrons above the solid-material surface. Equivalence principle (EP), i.e., inertia mass equals  the gravitational mass, is the basis for describing gravitational interaction. There are still no experimental evidences for EP violation~\cite{Rb,Exp}. However, String theories and some hypothetical experiments of gravity predict EP violation, see, e.g.,~\cite{EEP-RMP,EEP-LRR,Constraint}. Recently, scientists propose to test EP with unprecedented accuracy by the program of European Space Agency: the Space-Time Explorer and Quantum Equivalence Space Test mission (STE-QUEST)
~\cite{Space-EEP,Space-Theroy1,Space-Theroy2,
Space-Theroy3,Space-1015NJP,Space-1015CQG}. The advantages of space experiment are the long freely falling time of test particles and the small nongravity noises. Scientists are planning to measure the gravitational acceleration with a high accuracy of $10^{-15}g$~\cite{Space-1015NJP,Space-1015CQG}.

In theory, the calculations for AI's phase shifts are usually based on the uniform gravity and its first-order gradient~\cite{Chu-Metrologia,kasevich-APB,
Curvature}, the latter is known as a serious systematic effect in atomic tests of the universality of free fall~\cite{limitation1,limitation2,limitation3,limitation4}. In this article, we study quantum free fall with high-order gravity gradients. On one hand, the high-order gravity gradients may affect AI's phase shift in the future space laboratory. On the other hand, there has been considerable interest in quantum violation of the universality
of free fall~\cite{Greenberger0,Greenberger,UFF-PRD,UFF-CQG,free fall}. Greenberger~\cite{Greenberger0} first introduced this conception by using the gravitational Bohr atoms. The same as Coulomb force, Newtonian gravity is also the central force, so the orbital radius of a particle moving around Earth may be ``discrete", which depends on the mass of free fall and therefore contradicts the universality of free fall. However, this still needs a detailed study as the atomic initial state is unlike that of classical particles dropped from some local regimes.
In mathematics, it would be very difficult to construct a local initial wave packet in spatial 3-dimensions (wherein the test particles were dropped), via the superposition of `` hydrogen atomic eigenstates''.
Note that, even in classical mechanics, analytically solving the trajectory of free fall as a function of time is not trivial, which refers to the inverse Kepler problem~\cite{Kepler}.

In this work, we Taylor-expand gravitational potential to be the cubic form. With such nonlinear potential, the positional average of particles couples with its initial position and velocity uncertainties during the process of free falling. Then, there exists a phase shift in AI which depends on the velocity uncertainty of the incident atoms. The phase shift increases with the time of freely falling as $t^4$, and decreases with Earth's radius as $R^2$. Thus, this effect is negligible in the ground-based laboratory, because of large $R$ and short $t$. However, the present systematic effect may be not negligible in the future space experiment due to the gravity of the satellites whose sizes are far smaller than that of Earth, and very long freely falling time of cold atoms.

We further investigate the nonclassicality of free fall based on the Wigner equation which is known as an effective approach to study the quantum-to-classical transition~\cite{Wigner1932}. With the nonlinear Newtonian potential, the Wigner equation in phase space of position and velocity is different from the classical Liouville equation. There exists a quantum correction in the dynamical equation which depends on the mass of free fall. We propose an approach to solve the multidimensional Wigner equation within the lowest-order quantum correction, and give a numerical estimation on the magnitude of the observable nonclassical effect in real space.

This article is organized as follows. In Sec. II, by using the Newtonian equation (and Heisenberg equation) we calculate the time-dependent position (and positional operator) of a freely falling particle in the gravitational potential of cubic form.
The obtained solution of the positional operator takes the same form as that of the classical trajectory in time evolution. In Sec III., we study the path-dependent phase of AI with the cubic potential, and show that, there is a phase shift relating to the initial velocity uncertainty of incident atoms and may be detectable in the space laboratory. We also suggest using the nonuniform magnetic field to simulate gravitational field for demonstrating the predicted effects in the ground-based laboratory. In Sec. IV, we solve the gravitational Wigner equation and discuss the dynamical nonclassicality. Finally, we present our conclusions
in Sec. V.

\section{Position and positional operator}
\subsection{The classical one-dimensional motion}
In the region $x,y,z\ll R$ of the displaced Cartesian coordinates $(R+x,y,z)$, we expand the Newtonian potential as
\begin{equation}
\begin{aligned}
\phi(\textbf{r})&=-\frac{GM}{\sqrt{(R+x)^2+y^2+z^2}}\\
&\approx  gx-\frac{g }{R} x^2+\frac{g }{2R}(y^2+z^2)\\
& \,\,\,\,\,\,+\frac{g }{R^2} x^3-\frac{3g }{2R^2}(y^2+z^2)x+\cdots,
\end{aligned}
\end{equation}
by neglecting the terms of higher order than $(1/R)^2$ and the constant $-GM/R$, where $\textbf{r}=(x,y,z)$ and $g=GM/R^2$, with gravitational constant $G\approx6.67\times10^{-11}\,\text{N}\cdot\text{m}^2\cdot\text{kg}^{-2}$ and the mass $M$ of a spherical gravity source.
Following potential (1), the classical motion equation ${\rm d}^2x/{\rm d}t^2=-\partial_x\phi(\textbf{r})$ along the $x$-direction reads:
\begin{equation}
\begin{aligned}
\frac{{\rm d}^2x}{{\rm d} t^2}\approx- g\left[1-\frac{2 x}{R}+\frac{3 x^2}{R^2} -\frac{3(y^2+z^2)}{2R^2}\right]\,.
\end{aligned}
\end{equation}
In the ground-based laboratory, the freely falling time of atoms is ultimately limited by the sizes of  practical vacuum installation, for example, the freely falling of $2$ seconds needs a height about $20\,\text{m}$. However, the time can be very long in the microgravity system. For example, the biggest gravitational acceleration is $g_s=GM/R_s^2\approx6.67\times10^{-8}\,\text{m}/\text{s}^2$ at the surface of a sphere of mass $M=10^3\,\text{kg}$ and radius $R_s=1\,\text{m}$. The initial velocity of cold atoms is on the order of $v_0=10^{-3}\,\text{m}/\text{s}$, and therefore the characteristic length of motions is $x\sim v_0t_0+g_st_0^2/2\approx v_0t_0=10^{-2}\,\text{m}$
with the freely falling time $t_0=10\,\text{s}$. As a consequence, the acceleration contributed by the high-order gravity gradients is  $3gx^2/R^2\approx4\times10^{-12}\,\text{m}/\text{s}^2$, with $R=1.5\,\text{m}$. This acceleration seems small but may be detectable in the future space experiments with high precision.

Eliminating physical units in the left and right hands of Eq.~(2), all quantities in the equation can be temporarily regarded as dimensionless. In terms of small $1/R$, we expand the solution of Eq.~(2) as
\begin{equation}
x(t)=x_0(t)+\frac{1}{R} x_1(t)+\frac{1}{R^2}x_2(t).
\end{equation}
Inserting this series back into Eq.~(2) results in
\begin{equation}
\frac{{\rm d}^2 x_0}{{\rm d}t^2}
=-g\,,
\end{equation}
\begin{equation}
\frac{{\rm d}^2 x_1}{{\rm d}t^2}
=2gx_0\,,
\end{equation}
\begin{equation}
\frac{{\rm d}^2 x_2}{{\rm d}t^2}
=2gx_1-3gx_0^2+\frac{3}{2}g(y^2+z^2)\,.
\end{equation}
The solution of Eq.~(4) is well known:
\begin{equation}
x_0(t)=x_i+v_{xi}t-\frac{1}{2}gt^2\,,
\end{equation}
with $x_i$ and $v_{xi}$ being the initial position and initial velocity along the $x$-direction. Inserting Eq.~(7) into Eq.~(5), we have
\begin{equation}
x_1(t)=gt^2(x_i+\frac{1}{3}v_{xi}t-\frac{1}{12}gt^2)\,.
\end{equation}

For Eq.~(6), we must find the solutions of $y$- and $z$-directional motions. In those directions, $\partial_y\phi(\textbf{r})=0$ and $\partial_z\phi(\textbf{r})=0$, by neglecting the terms of $1/R$ in potential (1). Therefore, we can use the simplest forms $y(t)=y_i+v_{yi}t$ and $z(t)=z_i+v_{zi}t$ to solve Eq.~(6), because $x_2(t)/R^2$ is the considered highest-order correction of nonlinear gravity. Here, $(y_i,z_i)$ and $(v_{yi}, v_{zi})$ are the initial position and initial velocity in the $(y,z)$ plane. Then the solution of Eq.~(6) is given by
\begin{equation}
\begin{aligned}
x_2(t)=&\frac{5 g^2t^4}{12}x_i
+\frac{11 g^2t^5}{60}v_{xi}-\frac{11 g^3t^6}{360}\\
&
-\frac{3gt^2}{4}(2x_i^2-y_i^2-z_i^2)
-\frac{gt^4}{8}(2v_{xi}^2-v^2_{yi}-v^2_{zi})\\
&
-\frac{gt^3}{2}(2x_i v_{xi}-y_iv_{yi}-z_iv_{zi})\,.
\end{aligned}
\end{equation}
Finally, the solution of Eq.~(3) is shortly written as
\begin{equation}
\begin{aligned}
x(\textbf{r}_i,\textbf{v}_i,t)=&\alpha x_i+\beta v_{xi}
-\gamma\\
&
-\tilde{\alpha}(2x_i^2-y_i^2-z_i^2)-\tilde{\beta}
(2v_{xi}^2-v^2_{yi}-v^2_{zi})\\
&
-\tilde{\gamma}(2 x_iv_{xi}-y_iv_{yi}-z_iv_{zi})\,.
\end{aligned}
\end{equation}
with
\begin{equation}
\alpha=1+\frac{gt^2}{R}+\frac{5 g^2t^4}{12 R^2}\,,
\end{equation}
\begin{equation}
\beta=t(1+\frac{gt^2}{3R}+\frac{11 g^2t^4}{60R^2})\,,
\end{equation}
\begin{equation}
\gamma=\frac{1}{2}gt^2(1+\frac{gt^2}{6R}+\frac{11 g^2t^4}{180R^2})\,,
\end{equation}
\begin{equation}
\tilde{\alpha}=\frac{3 gt^2}{4R^2}\,,\,\,\,\,\,\,
\tilde{\beta}=\frac{gt^4}{8R^2}\,,\,\,\,\,\,\,
\tilde{\gamma}=\frac{gt^3}{2R^2}\,.
\end{equation}
\begin{figure}[tbp]
\includegraphics[width=9cm]{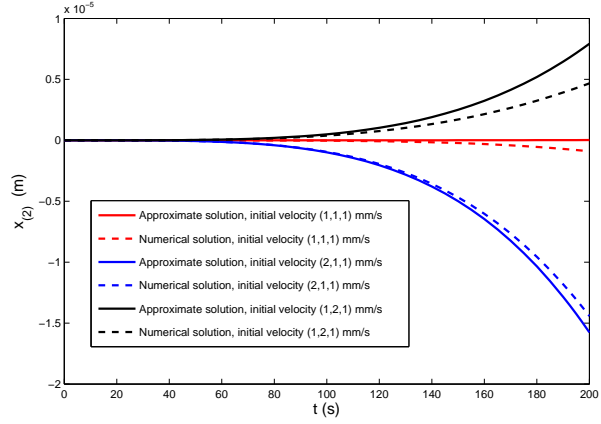}
\caption{The solutions of $x_{(2)}$, assuming a free fall was dropped from the initial position $(R+x,y,z)=(1.5+0,0,0)\,\text{m}$, near a sphere of mass $M=10^3\,\text{kg}$ and radius $R_s=1\,\text{m}$. The solid lines are obtained from the analytical solution~(10) and the dashed lines are the results from the numerical solution of Newtonian equation~(15). Three initial velocities are considered: $(1,1,1)\,\text{mm}/\text{s}$, $(2,1,1)\,\text{mm}/\text{s}$, and $(1,2,1)\,\text{mm}/\text{s}$.}
\end{figure}
The above parameters are rather tedious, but can be verified by numerically solving the exact Newtonian equation (three coupled differential equations):
\begin{equation}
\begin{aligned}
\frac{{\rm d}^2\textbf{r}}{{\rm d}t^2}
=\nabla_r\frac{GM}{\sqrt{(R+x)^2+y^2+z^2}}\,,
\end{aligned}
\end{equation}
with $\nabla_r=(\partial_x,\partial_y,\partial_z)$, whose solution in the $x$-direction is denoted by $x_{\rm num}$. Correspondingly, we temporarily denote the approximate solution (10) as $x_{\rm app}$, wherein the second-order correction $x_{(2)}=x_{\rm app}-x_0-(x_1/R)=x_2/R^2$ is the feature in the present work. So, one may compare it with the numerical counterpart $x_{\rm num}-x_0-(x_1/R)$, based on the given $x_0$ and $x_1$ by Eqs.~(7) and (8) respectively. Using our estimated parameters in the paragraph below Eq.~(2), we found that the analytical solution in short duration is in good agreement with the numerical one, see Fig.~1.

\subsection{Heisenberg equation for positional operator}
We now generalize the above perturbation theory into quantum mechanics for computing the time-dependent positional operator of free fall.
The quantized Hamiltonian reads $\hat{H}=\hat{\textbf{p}}^2/(2m)+m\phi(\hat{\textbf{r}})$,
where the momentum operator $\hat{\textbf{p}}=(\hat{p}_x,\hat{p}_y,\hat{p}_z)$ and position operator $\hat{\textbf{r}}=(\hat{x},\hat{y},\hat{z})$ obey the canonical quantization.
According to the Schr\"{o}dinger equation, the time-dependent state can be written as $|\psi\rangle=\exp(-it\hat{H}/\hbar)|i\rangle$ with an initial state  $|i\rangle$. Therefore, the time evolution of any observable quantity $\langle i|\hat{O}(t)|i\rangle$ is determined by
\begin{equation}
\hat{O}(t)=e^{\frac{it}{\hbar}\hat{H}}\hat{O}e^{\frac{-it}{\hbar}\hat{H}}\,.
\end{equation}
The Heisenberg equation of $\hat{O}(t)$ is
\begin{equation}
\frac{{\rm d}\hat{O}(t)}{{\rm d}t}=\frac{i}{\hbar}
e^{\frac{it}{\hbar}\hat{H}}[\hat{H},\hat{O}]e^{\frac{-it}{\hbar}\hat{H}}
=\frac{i}{\hbar}
[\hat{H},\hat{O}](t)\,.
\end{equation}

According to canonical quantization $[\hat{x},\hat{p}_x]=i\hbar$, we have $[\hat{p}_x^2,\hat{x}]=-i2\hbar\hat{p}_x$ and $[\phi(\hat{\textbf{r}}),\hat{p}_x]=i\hbar\partial_{\hat{x}}\phi(\hat{\textbf{r}})$ along the $x$-direction, and consequently the quantized canonical equations: ${\rm d}\hat{x}(t)/{\rm d}t=\hat{p}_x(t)/m$, ${\rm d}\hat{p}_x(t)/{\rm d}t=-m[\partial_{\hat{x}}\phi(\hat{\textbf{r}})](t)$, and ${\rm d}^2\hat{x}(t)/{\rm d} t^2=-[\partial_{\hat{x}}\phi(\hat{\textbf{r}})](t)$. These equations can be easily generalized into three spatial dimensions, ${\rm d}\hat{\textbf{r}}(t)/{\rm d}t=\hat{\textbf{p}}(t)/m$, ${\rm d}\hat{\textbf{p}}(t)/{\rm d}t=-m[\nabla_{\hat{r}}\phi(\hat{\textbf{r}})](t)$, and ${\rm d}^2\hat{\textbf{r}}(t)/{\rm d} t^2=-[\nabla_{\hat{r}}\phi(\hat{\textbf{r}})](t)$, with $\nabla_{\hat{r}}=(\partial_{\hat{x}},\partial_{\hat{y}},\partial_{\hat{z}})$.
Consequently, using the approximate potential (1), the Heisenberg equation of $\hat{x}(t)$ reads
\begin{equation}
\begin{aligned}
\frac{{\rm d}^2\hat{x}(t)}{{\rm d} t^2}\approx-g+\frac{2g \hat{x}(t)}{R}-\frac{3g[\hat{x}(t)]^2}{R^2} +\frac{3 g[\hat{y}(t)]^2+3g[\hat{z}(t)]^2}{2R^2}\,,
\end{aligned}
\end{equation}
which takes the same form to the classical Eq.~(2). Thus, the previous approach for solving Eq.~(2) is also effective for the present equation.

We also expand the Heisenberg operator as
\begin{equation}
\hat{x}(t)=\hat{x}_0(t)+\frac{1}{R} \hat{x}_1(t)+\frac{1}{R^2}\hat{x}_2(t).
\end{equation}
Inserting this series into Eq.~(18), and using the approximate solutions $\hat{y}(t)=\hat{y}+\hat{v}_yt$, $\hat{z}(t)=\hat{z}+\hat{v}_zt$ in $y$- and $z$- directions, we find
\begin{equation}
\hat{x}_0(t)=\hat{x}+\hat{v}_xt-\frac{1}{2}gt^2\,,
\end{equation}
\begin{equation}
\hat{x}_1(t)=g t^2(\hat{x}+\frac{1}{3}\hat{v}_xt-\frac{1}{12}gt^2)\,,
\end{equation}
and
\begin{equation}
\begin{aligned}
\hat{x}_2(t)=&\frac{5 g^2t^4}{12}\hat{x}
+\frac{11 g^2t^5}{60}\hat{v}_x-\frac{11 g^3t^6}{360}\\
&
-\frac{3 gt^2}{4}(2\hat{x}^2-\hat{y}^2-\hat{z}^2)
-\frac{gt^4}{8}(2\hat{v}_x^2-\hat{v}^2_{y}-\hat{v}^2_{z})\\
&
-\frac{gt^3}{2}(\hat{x}\hat{v}_x+\hat{v}_x\hat{x})
+\frac{g t^3}{4}(\hat{y}\hat{v}_{y}
+\hat{v}_{y}\hat{y}+\hat{z}\hat{v}_{z}+\hat{v}_{z}\hat{z})\,,
\end{aligned}
\end{equation}
where $\hat{\textbf{v}}=\hat{\textbf{p}}/m=(\hat{p}_x,\hat{p}_y,\hat{p}_z)/m$ is the velocity operator. Such a definition is nothing but shows very obvious  correspondence between classical and quantum mechanics. Indeed, if one replaces $\hat{\textbf{r}}=(\hat{x},\hat{y},\hat{z})$ and $\hat{\textbf{v}}=(\hat{v}_x,\hat{v}_y,\hat{v}_z)$ by the classical initial position $\textbf{r}_i=(x_i,y_i,z_i)$ and initial velocity $\textbf{v}_i=(v_{xi},v_{yi},v_{zi})$ respectively, Eqs.~(20)-(22) will exactly reduce to the classical solutions (7)-(9).
Inversely, one can directly quantize the classical solutions by $\textbf{r}_i\rightarrow\hat{\textbf{r}}$ and $\textbf{v}_i\rightarrow\hat{\textbf{v}}$. In such a way, the product between position and velocity in classical formulas should be rewritten as $\textbf{r}_i\cdot \textbf{v}_i=(\textbf{r}_i\cdot \textbf{v}_i+\textbf{v}_i\cdot\textbf{r}_i)/2$.

There are two purposes for listing out Eqs.~(19)-(22). First, they are needed in Sec. III for computing AI's phase in the orders of $(1/R)^0$, $(1/R)^1$ and $(1/R)^2$. Second, they show very obvious correspondence between classical and quantum mechanics. Following these equations, the positional average is given as
\begin{equation}
\begin{aligned}
\langle i|\hat{x}(t)|i\rangle=&\alpha\langle i|\hat{x}|i\rangle
+\beta\langle i|\hat{v}_x|i\rangle
-\gamma\\
&
-\tilde{\alpha}\langle i|(2\hat{x}^2-\hat{y}^2-\hat{z}^2)|i\rangle
-\tilde{\beta}
\langle i|(2\hat{v}_x^2-\hat{v}_y^2-\hat{v}_z^2)|i\rangle\\
&
-\tilde{\gamma}\langle i|(\hat{v}_x\hat{x}+\hat{x}\hat{v}_x)|i\rangle
+\frac{\tilde{\gamma}}{2}\langle i| (\hat{y}\hat{v}_y+\hat{v}_y \hat{y}+\hat{z}\hat{v}_z+\hat{v}_z \hat{z})|i\rangle\,.
\end{aligned}
\end{equation}
It depends on initial uncertainties of position and velocity, and the position-velocity correlation of the prepared particles ensemble. In any case,
compared to the classical statistics based on solution (10), the time-dependent parameters $\alpha$, $\beta$, etc. are not changed. Thus, if the initial state is given by a classical interpretation, for example the Gaussian distribution in Wigner representation~\cite{PRDG}, then the above time-dependent expectation value is also classical.
In Sec. IV, we will continue to discuss this issue based on the Wigner equation.

\section{Atom interferometer}
In experiments, the gravitational acceleration of atom is usually measured by AI. In such a device, the atom is manipulated by laser pulses and moves simultaneously along two separate paths to arrive at the detector. In the uniform gravitational field, the path-dependent interferometric phase is independent on the initial position, initial velocity, and the mass of atom. This allows scientists to test EP with very high precision based on the Hamiltonian $\hat{H}_0=\hat{\textbf{p}}^2/(2m)+mg\hat{x}$, where the inertia mass is equal to the gravitational mass, $m_{\rm I}=m_{\rm G}=m$.
However, the interferometric phase will become very complicated when the nonlinear gravity is present. In principle, the atomic initial motions cannot be completely eliminated. Usually, the mathematical derivations for AI's interferometric phase are based on the Feynman path integral~\cite{tutorial, wave-packet,CQG}. In this work, we derive the interferometric phase by using the well-known Zassenhaus formula:
\begin{equation}
e^{\hat{A}+\hat{B}}=e^{\hat{A}}e^{\hat{B}}e^{-\frac{1}{2}[\hat{A},\hat{B}]}
e^{\frac{1}{6}[\hat{A},[\hat{A},\hat{B}]]+\frac{1}{3}[\hat{B},[\hat{A},\hat{B}]]}\cdots\,.
\end{equation}
It includes infinite nested-commutator of operators $\hat{A}$ and $\hat{B}$, but can be exactly (or approximately) solved within some special cases, for example, $[\hat{A},[\hat{A},\hat{B}]]$ and $[\hat{B},[\hat{A},\hat{B}]]$ equal to the $c$-numbers. In the following, we also frequently use the Baker-Campbell-Hausdorff (BCH) formula~\cite{Z-2012,Z-1967},
\begin{equation}
e^{\hat{A}}\hat{B}e^{-\hat{A}}=\hat{B}+[\hat{A},\hat{B}]
+\frac{1}{2!}[\hat{A},[\hat{A},\hat{B}]]
+\frac{1}{3!}[\hat{A}, [\hat{A},[\hat{A},\hat{B}]]]+\cdots\,.
\end{equation}

\subsection{Zassenhaus approach for atom interferometer}
\begin{figure}[tbp]
\includegraphics[width=8cm]{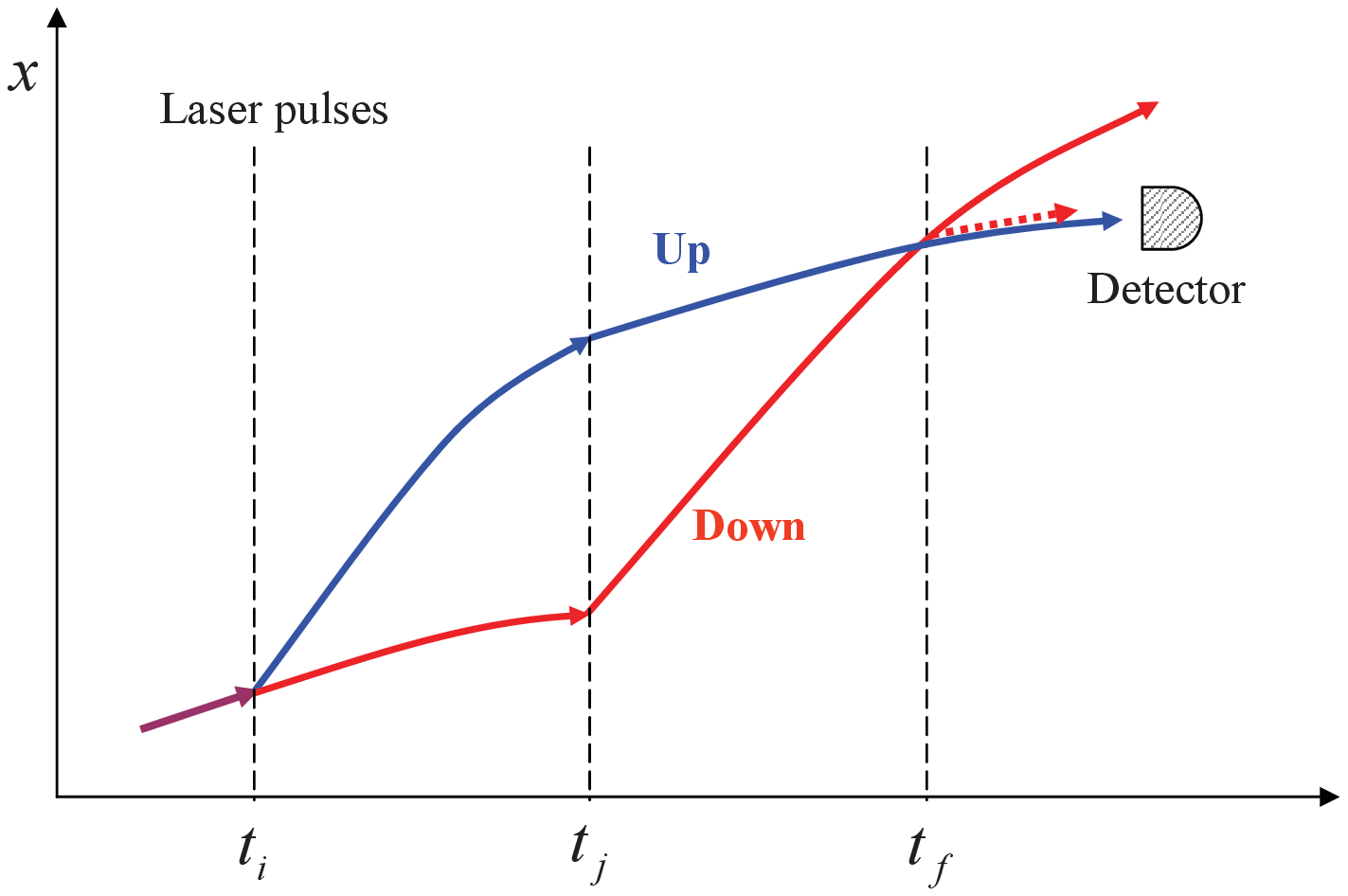}
\caption{Sketch for atom interferometer. Three short laser pulses are applied to split and recombine the paths of the traveling atom. The durations for freely falling in each stage are equal, i.e., $t_j-t_i=t_f-t_j=t$. Due to the second-order gravity gradient, the interferometric phase can be shifted by changing the initial momentum uncertainty of the incident atom.}
\end{figure}
In a typical AI, see Fig.~2, three laser pulses are applied to probabilistically change atomic momentum by $\hbar \vec{k}$, with $\vec{k}$ being the wave vector of the applied laser pulses. For simplicity, we assume that the $y$- and $z$-directional components of $\vec{k}$ are negligible. In Fig.~2, the atom moving along the ``up''-path
undergos four changes: (a) momentum instantaneously increasing due to the laser scattering at time $t_i$, (b) freely falling with a duration of $t=t_j-t_i$, (c) momentum suddenly decreasing by the second laser pulse at $t_j$, and (d) the freely falling in stage $t=t_f-t_j$. So, the state evolution of the ``up''-atom is written as
\begin{equation}
\begin{aligned}
|\uparrow\rangle=\hat{U}(t)e^{-ik\hat{x}}\hat{U}(t)e^{ik\hat{x}}|i\rangle\,.
\end{aligned}
\end{equation}
The operator $\exp(\pm ik\hat{x})$ describes the momentum changes due to the laser scattering, and $\hat{U}(t)=\exp(-it\hat{H}/\hbar)$ is the evolution operator of free fall. There also exists a probability that the atom moves along the ``down''-path, i.e., the laser induced momentum increasing and decreasing occur at the times $t_j$ and $t_f$, respectively. Thus, we write the state evolution for the ``down''-path atom as
\begin{equation}
\begin{aligned}
|\downarrow\rangle=e^{-ik\hat{x}}\hat{U}(t)e^{ik\hat{x}}\hat{U}(t)|i\rangle\,.
\end{aligned}
\end{equation}
The $|\uparrow\rangle$-atom and $|\downarrow\rangle$-atom move both toward to the detector, resulting in an interferometric signal of $\langle \downarrow|\uparrow\rangle+\text{c.c.}$, where
\begin{equation}
\begin{aligned}
\langle \downarrow|\uparrow\rangle=\langle i|\hat{U}^\dagger(t) e^{-ik\hat{x}}\hat{U}^\dagger(t) e^{ik\hat{x}}
\hat{U}(t)e^{-ik\hat{x}}\hat{U}(t)e^{ik\hat{x}}|i\rangle\,.
\end{aligned}
\end{equation}

Using the relation $\hat{U}(t)\hat{U}^\dagger(t)=\hat{U}^\dagger(t)\hat{U}(t)=1$ and the solution $\hat{U}^\dagger(t)\hat{x}\hat{U}(t)=\hat{x}(t)$ in Sec. II, we further write (28) as
\begin{equation}
\begin{aligned}
\langle \downarrow|\uparrow\rangle=\langle i|e^{-ik\hat{x}(t)}e^{ik\hat{x}(2t)}e^{-ik\hat{x}(t)}e^{ik\hat{x}}|i\rangle\,.
\end{aligned}
\end{equation}
In terms of $1/R$, the positional operator was perturbatively expanded as $\hat{x}(t)=\hat{x}_0(t)+\hat{x}_1(t)/R+\hat{x}_2(t)/R^2$.
As a consequence, the exponent is expanded as
\begin{equation}
\begin{aligned}
e^{-ik\hat{x}(t)}\approx e^{-ik\hat{x}_0(t)}\hat{L}(t,k,\hat{\textbf{r}},\hat{\textbf{v}})
\end{aligned}
\end{equation}
by using the Zassenhaus formula. Here,
\begin{equation}
\begin{aligned}
\hat{L}(t,k,\hat{\textbf{r}},\hat{\textbf{v}})
=& e^{-i\frac{k\hat{x}_1}{R}}
e^{\frac{k^2}{2R}[\hat{x}_0,\hat{x}_1]}\\
&\times e^{-i\frac{k\hat{x}_2}{R^2}}
e^{\frac{k^2}{2R^2}[\hat{x}_0,\hat{x}_2]}\\
&\times
e^{i\frac{k^3}{6R^2}[\hat{x}_0,[\hat{x}_0,\hat{x}_2]]}\,,
\end{aligned}
\end{equation}
with
\begin{equation}
\begin{aligned}
&[\hat{x}_0,\hat{x}_1]=-\frac{2gt^3}{3}\frac{i\hbar}{m}\,,\\
&[\hat{x}_0,\hat{x}_2]=
(2gt^3\hat{x}+\frac{gt^4}{2}\hat{v}_x-\frac{7g^2t^5}{30})\frac{i\hbar}{m}\,,\\
&[\hat{x}_0,[\hat{x}_0,\hat{x}_2]]
=\frac{3gt^4}{2}\frac{\hbar^2}{m^2}\,.
\end{aligned}
\end{equation}
In the above Zassenhaus formula, the terms with the higher orders than $1/R^2$ have been neglected, as well as that in Sec. II.

Using formula (30), the exponents product in formula (29) is written as
\begin{equation}
\begin{aligned}
\hat{K}&=
e^{-ik\hat{x}(t)}e^{ik\hat{x}(2t)}e^{-ik\hat{x}(t)}e^{ik\hat{x}}\\
&
=e^{-ik\hat{x}_0(t)}\hat{L}(t,k,\hat{\textbf{r}},\hat{\textbf{v}})
e^{ik\hat{x}_0(2t)}\hat{L}(2t,-k,\hat{\textbf{r}},\hat{\textbf{v}})
e^{-ik\hat{x}_0(t)}\hat{L}(t,k,\hat{\textbf{r}},\hat{\textbf{v}})e^{ik\hat{x}}\\
&
=\hat{L}(t,k,\hat{\textbf{r}}-\hat{\textbf{r}}_d,\hat{\textbf{v}}+\hat{\textbf{v}}_d)
\hat{L}(2t,-k,\hat{\textbf{r}}+\hat{\textbf{r}}_d,\hat{\textbf{v}})
\hat{L}(t,k,\hat{\textbf{r}},\hat{\textbf{v}}+\hat{\textbf{v}}_d)
e^{-igkt^2}\,.
\end{aligned}
\end{equation}
In the last line of the above equation, we have used the identities:
\begin{equation}
\begin{aligned}
e^{-ik\hat{x}_0(t)}\hat{\textbf{v}}e^{ik\hat{x}_0(t)}
=\hat{\textbf{v}}+\hat{\textbf{v}}_d
\,,\,\,\,\,\,\,\,\hat{\textbf{v}}_d=\frac{\hbar k}{m}(1,0,0)\,,
\end{aligned}
\end{equation}
\begin{equation}
\begin{aligned}
e^{-ik\hat{x}_0(t)}\hat{\textbf{r}}e^{ik\hat{x}_0(t)}
=\hat{\textbf{r}}-\hat{\textbf{r}}_d
\,,\,\,\,\,\,\,\hat{\textbf{r}}_d=t\hat{\textbf{v}}_d\,.
\end{aligned}
\end{equation}

The first result from Eq.~(33) is $\theta_0=gkt^2$, the well-known phase of AI generated by the uniform gravity. If one neglects the corrections from the order $1/R^2$, Eq.~(31) reduces to
\begin{equation}
\begin{aligned}
\hat{L}_1&=e^{-i\frac{\theta_0}{R}(\hat{x}+\frac{1}{3}\hat{v}_xt)}
e^{i\frac{\theta_0gt^2}{12R}}
e^{-i\frac{\theta_0\hbar kt}{3mR}}\,.
\end{aligned}
\end{equation}
As a consequence, the operator $\hat{K}$ reduces to
\begin{equation}
\begin{aligned}
\hat{K}_{1}&=
e^{-i\frac{\theta_0}{R}(\hat{x}+\frac{\hat{v}_xt}{3})}
e^{i \frac{4\theta_0}{R}(\hat{x}+\frac{2\hat{v}_xt}{3})}
e^{-i\frac{\theta_0}{R}(\hat{x}+\frac{\hat{v}_xt}{3})}e^{-i(\theta'+\theta_0)}\\
&= e^{i\frac{2\theta_0}{R}(\hat{x}+\hat{v}_xt)}e^{-i(\theta'+\theta_0)}\,,
\end{aligned}
\end{equation}
with
\begin{equation}
\begin{aligned}
\theta'=\frac{7\theta_0gt^2}{6R}-\frac{\theta_0\hbar kt}{mR}\,.
\end{aligned}
\end{equation}
The phase $\theta'$ is due to the gravity gradient $2g/R$, and independent on the initial state $|i\rangle$ of the atom. In Refs.~\cite{kasevich-APB,Curvature}, there still exists a phase relating to the initial velocity of the atom. For that, we Taylor expand  $\exp[i2\theta_0(\hat{x}+\hat{v}_xt)/R]$, and consequently,
\begin{equation}
\begin{aligned}
\langle i|\hat{K}_1|i\rangle\approx\left[1+i\frac{2\theta_0}{R}(\langle i|\hat{x}|i\rangle+t\langle i|\hat{v}_x|i\rangle)\right]
e^{-i(\theta'+\theta_0)}\,.
\end{aligned}
\end{equation}
Indeed, there is a phase $\theta_{ v_x}=2\theta_0t\langle i|\hat{v}_x|i\rangle/R$ referring to the atomic initial central velocity $v_{xi}=\langle i|\hat{v}_x|i\rangle$ along the $x$-direction.

\subsection{Phase shift due to the initial velocity uncertainty}
In terms of $1/R$ and $1/R^2$, we write formula (31) as $\hat{L}=\hat{L}_1\hat{L}_2$, with
\begin{equation}
\begin{aligned}
\hat{L}_2(t,k,\hat{\textbf{r}},\hat{\textbf{v}})= e^{-i\frac{k\hat{x}_2}{R^2}}
e^{\frac{k^2}{2R^2}[\hat{x}_0,\hat{x}_2]}
e^{i\frac{k^3}{6R^2}[\hat{x}_0,[\hat{x}_0,\hat{x}_2]]}\,.
\end{aligned}
\end{equation}
Note that, $[\hat{L}_1,\hat{L}_2]\approx0$ and $[\hat{L}_2(t,k),\hat{L}_2(2t,-k)]\approx0$, because the outcomes of such commutators are on the order of $1/R^3$.
As a consequence, $\hat{K}\approx\hat{K}_1\hat{K}_2$, with
\begin{equation}
\begin{aligned}
\hat{K}_2&=\hat{L}_2(t,k,\hat{\textbf{r}}-\hat{\textbf{r}}_d,
\hat{\textbf{v}}+\hat{\textbf{v}}_d)
\hat{L}_2(2t,-k,\hat{\textbf{r}}+\hat{\textbf{r}}_d,\hat{\textbf{v}})
\hat{L}_2(t,k,\hat{\textbf{r}},\hat{\textbf{v}}+\hat{\textbf{v}}_d)\\
&\approx e^{i\frac{k}{R^2}[\hat{x}_{2}(2t)-2\hat{x}_{2}(t)]}
e^{i(\mu\hat{x}+\nu\hat{v}_xt)}e^{i\theta''}\,,
\end{aligned}
\end{equation}
Above, $\theta''$ is a real number and independent on the initial state of the atom, as well as the standard phase shift $\theta_0$ and $\theta'$ from the first-order gravity gradient. The $\mu$ and $\nu$ are the coefficients for the linear operators $\hat{x}$ and $\hat{v}_x$ in  $\exp[i(\mu\hat{x}+\nu\hat{v}_xt)]$. This term generates a phase that depends on the central position and the central velocity of incident atoms, see Eq.~(39).

In $\hat{K}_2$, the operator $\hat{x}_{2}(t)$ has been already solved in Sec. II, which contains the quadratic operators, such as $\hat{v}_x^2$. Thus, Taylor expanding the exponential operator in $\langle i|\hat{K}_1\hat{K}_2|i\rangle$, one can find that there exists a phase depending on the velocity uncertainty of incident state $|i\rangle$. This is one of the new systematic effects different from that of the first-order gravity gradient. For demonstrating such effect, one can apply an additional manipulation (before the standard AI) to change the initial momentum uncertainty of incident atoms but not change the central momentum and central position of atomic wave packet. This manipulation can be realized by a pulse of standing-wave light which generates the atomic Kapitza-Dirac scattering~\cite{KD1,KD2,KD3}.
The phase shift due to such an additional laser pulse along the $x$-direction is
\begin{equation}
\begin{aligned}
\theta_{ v_x^2}=
\frac{7 gkt^4}{2R^2}\langle i|\hat{v}_x^2|i\rangle\,.
\end{aligned}
\end{equation}

Numerically, assuming $\sqrt{\langle i|\hat{v}_x^2|i\rangle}=0.1\hbar k/m\approx10^{-3}\,\,\text{m}/\text{s}$ with $k=2\pi/(500\,\text{nm})$ and $m=10^{-25}$~kg, we have $\theta_{ v_x^2}\approx5.9\pi \times10^{-12}$ with $t=1$~s. Here, $R\approx 6.4\times 10^6$~m and  $g=GM/R^2\approx9.8~\text{m}/\text{s}^2$ are respectively Earth's radius and the gravitational acceleration at Earth's surface. Compared to the standard phase shift $\theta_0$ and the $\theta_{v_x}$ of gravity gradient, the present phase shift is very small,
\begin{equation}
\begin{aligned}
\frac{\theta_{v_x^2}}{\theta_{0}}=
\frac{7 t^2\langle i|\hat{v}_x^2|i\rangle}{2R^2}\ll\frac{\theta_{v_x^2}}{\theta_{v_x}}
=\frac{7t\langle i|\hat{v}_x^2|i\rangle}{4R\langle i|\hat{v}_x|i\rangle}\ll1\,,
\end{aligned}
\end{equation}
because of large $R$ and short $t$ in the ground-based laboratory.  However, the situation would be different in microgravity environments, for example, the  gravity of a satellite in the space laboratory. Using the data $M=10^{3}\,\text{kg}$, $R=1.5\,\text{m}$, and $t=10\,\text{s}$ in the paragraph below Eq.~(2), we have $g=GM/R^2\approx3\times 10^{-8}\,\text{m}/\text{s}^2$,  $\theta_{0}\approx12\pi$, and $\theta_{v_x^2}\approx3.2\pi\times10^{-3}$. The phase shifts get smaller linearly with decreasing $M$, for example, $\theta_{0}\approx1.2\pi$ and $\theta_{v_x^2}\approx3.2\pi\times10^{-4}$ with $M=100\,\text{kg}$. The background noises in space laboratory, such as the blackbody radiation and the seismic noise (the vibration of experimental platform), are much smaller than those in the ground-based laboratory. Therefore, the atomic coherent time $t=10\,\text{s}$ or longer is possible~\cite{Space-1015NJP,Space-1015CQG}.

\subsection{Gravity simulation}
For experimentally demonstrating the above effects in the ground-based laboratory, one can use electromagnetic force to simulate gravity. Consider a magnetic field, $\vec{B}=\vec{e}_{\rm n}\mu_0I/[2\pi (R+x)]$, that is generated by a dc current $\vec{I}$ along the vertical direction. Here, $\mu_0$ is the permeability of vacuum, $\vec{e}_{\rm n}$ is the unit vector normal to the plane of $I$ and $R+x$. Here, $R+x$ is the distance between the atom and current, and $x$ is the dynamical position of the atom. Within the limitation $x\ll R$, the potential energy of atom in the nonuniform magnetic field can be expanded as
\begin{equation}
\begin{aligned}
\phi_b=-\frac{\mu_0IM_b}{2\pi (R+x)}\approx mg_bx-\frac{mg_b}{R} x^2+\frac{mg_b}{R^2} x^3\,,
\end{aligned}
\end{equation}
with $M_b$ being the effective magnetic moment of the atom, and acceleration $g_b=\mu_0IM_b/(2\pi R^2m)$. The present equation is very like the gravitational (1). The advantage is that the value of $g_b/R^2$ can be much larger than the gravitational counterpart.
Numerically, supposing $R=0.1$~m, $g_b=0.1~\text{m}/\text{s}^2$, and $t=0.1$~s, we have $\theta_{0}\approx4\pi\times10^3$ and $\theta_{ v_x^2}\approx0.025\pi$, with the same values of $m$, $k$, and $\langle i|\hat{v}_x^2|i\rangle$ as before.

\section{Nonclassicality in terms of Wigner equation}
In Sec.~III, quantum interference was obtained by applying several laser pulses to create a superposition state of atomic momentum via the help of atomic internal bound states. One might ask whether gravity itself generates the genuine nonclassicality regardless of laser manipulations. In this section, we discuss this issue based on the Wigner equation, which was introduced first by E. Wigner in 1932~\cite{Wigner1932}. This equation is equivalent to the Schr\"{o}dinger equation for the dynamical evolution of a pure state, and thus sometimes called
Wigner equivalent formalism~\cite{Heller}. At the same time, this equation also works for the mix state~\cite{MonteCarlo,LB-PRE}. The Wigner equation looks similar to the Newtonian equation in phase space and allows us to make an intuitive comparison between classical and quantum mechanics~\cite{AmJP}. In terms of the Wigner equation, the quantum correction is on the order of $\hbar^2/m^2$, and the arose gravitational quantum effects are too weak to be experimentally demonstrated so far.

\subsection{Wigner function}
Without generality, we start with the Wigner function in spatial 3-dimensions. It is defined as~\cite{Wigner1932}
\begin{equation}
W(\textbf{r},\textbf{p},t)
=\frac{1}{(2\pi\hbar)^3}\int_{-\infty}^{\infty}
e^{i\textbf{p}\cdot \textbf{r}'/\hbar}\psi^*(\textbf{r}+\textbf{r}'/2,t)\psi(\textbf{r}-\textbf{r}'/2,t) {\rm d}^3r'\,,
\end{equation}
with $\psi(\textbf{r},t)$
being the time-dependent state of a quantum system, and where $\textbf{r}=(x,y,z)$ and $\textbf{p}=(p_x,p_y,p_z)$. Following this definition, any observable quantity of the system can be formally written as~\cite{AmJP}:
\begin{equation}
\langle \psi|\hat{O}|\psi\rangle=\int_{-\infty}^{\infty} \widetilde{O}W(\textbf{r},\textbf{p},t) {\rm d}^3r {\rm d}^3p\,,
\end{equation}
with
\begin{equation}
\widetilde{O}=\int_{-\infty}^{\infty}e^{-i\textbf{p}\cdot \textbf{r}'/\hbar}\langle \textbf{r}+\textbf{r}'/2|\hat{O}|\textbf{r}-\textbf{r}'/2\rangle {\rm d}^3r'
\end{equation}
being the so-called Weyl transform of a Hermitian operator $\hat{O}$. In this representation, the observable quantity $\widetilde{O}$ is a function of $\textbf{r}$ and $\textbf{p}$, not the function of operators $\hat{\textbf{r}}$ and $\hat{\textbf{p}}$.

It has been well known: $\int_{-\infty}^{\infty}W(\textbf{r},\textbf{p},t) {\rm d}^3p{\rm d}^3r=1$, $|\psi(\textbf{r},t)|^2=\int_{-\infty}^{\infty}W(\textbf{r},\textbf{p},t) {\rm d}^3p$, and $|\varphi(\textbf{p},t)|^2=\int_{-\infty}^{\infty}W(\textbf{r},\textbf{p},t) {\rm d}^3r$ with $\varphi(\textbf{p},t)$ being the wave function in momentum Hilbert space of state $\psi(\textbf{r},t)$. Moreover, if $\hat{O}$ is purely a function of $\hat{x}$ or $\hat{p}_x$, then its
Weyl transform is just the original function~\cite{AmJP}, i.e, $\widetilde{x^n}=x^n$ and $\widetilde{p_x^n}=p_x^n$, with $n=0,1,2,\cdots$. Consequently, $\langle x\rangle=\int_{-\infty}^{\infty}xW(\textbf{r},\textbf{p},t){\rm d}^3r{\rm d}^3p$, $\langle x^2\rangle=\int_{-\infty}^{\infty}x^2
W(\textbf{r},\textbf{p},t){\rm d}^3r{\rm d}^3p$, etc. The $y$- and $z$-directional formulas take the similar forms. Because these representations are the same as the classical statistics in phase space, the Wigner function is also called Wigner quasiprobability distribution.
Note that, the Weyl transform of $\hat{x}\hat{p}_x+\hat{p}_x\hat{x}$ is also its original function, i.e., $\widetilde{xp_x}+\widetilde{p_xx}=2xp_x$. Hence, the position or momentum average with the given Heisenberg operator (19) in Sec. II can be regarded as a classical measurement to the initial Wigner function $W(\textbf{r},\textbf{p},0)$. However, the results would be different for some other measurable quantities, such as the distribution function $|\psi(\textbf{r},t)|^2$ in real space. Thus, it is necessary to study the time-dependent Wigner function $W(\textbf{r},\textbf{p},t)$.

\subsection{Wigner equation with lowest-order quantum correction}
The time evolution of wave function $\psi(\textbf{r},t)$ obeys Schr\"{o}dinger equation, and thus one can establish a dynamical equation for the Wigner function, with definition (45) and the Hamiltonian $\hat{H}=\hat{\textbf{p}}^2/(2m)+V(\textbf{r})$. Solving the original $(6+1)$-dimensional Wigner equation is a huge challenge~\cite{MonteCarlo,LB-PRE}, especially with the present central force problem. Nevertheless, it is resolvable within the classical limitation $\hbar^2\rightarrow0$~\cite{PhysReport,Lee-Scully,PRL-Martens}.
In such a limitation, Wigner equation reads~\cite{MonteCarlo}
\begin{equation}
\partial_tW(\textbf{r},\textbf{p},t)
+\frac{\textbf{p}}{m}\cdot\nabla_r W(\textbf{r},\textbf{p},t)-\nabla_r V(\textbf{r})\cdot\nabla_pW(\textbf{r},\textbf{p},t)=Q\,,
\end{equation}
with
\begin{equation}
Q=\frac{-\hbar^{2}}{24}\nabla^3_{r}V(\textbf{r})\cdot\nabla^3_pW(\textbf{r},\textbf{p},t)
+\mathcal{O}(\hbar^2)\,,
\end{equation}
where $\nabla_r=(\partial_x,\partial_y,\partial_z)$ and $\nabla_p=(\partial_{p_x},\partial_{p_y},\partial_{p_z})$.
The $\hbar^2$-dependent $Q$ can be regarded as a quantum mechanical correction to the dynamical system, because if $Q=0$, Eq.~(48) is exactly the classical Liouville equation. In Eq.~(49), the high orders of $\hbar^2$ have been neglected. This approximation was numerically verified by many previous references for the one-dimensional questions, see, e.g.,~\cite{PhysReport,Lee-Scully,PRL-Martens}. In terms of velocity $\textbf{v}=\textbf{p}/m$ and acceleration $\textbf{g}(\textbf{r})=-\nabla_rV(\textbf{r})/m=-\nabla_r\phi(\textbf{r})$, the Wigner equation can be further written as
\begin{equation}
\partial_tf(\textbf{r},\textbf{v},t)+\textbf{v}\cdot\nabla_r f(\textbf{r},\textbf{v},t)+\textbf{g}(\textbf{r})\cdot
\nabla_vf(\textbf{r},\textbf{v},t)=Q\,,
\end{equation}
with $\nabla_v=(\partial_{v_x},\partial_{v_y},\partial_{v_z})$.
Using our approximate potential (1),
the quantum correction reads
\begin{equation}
\begin{aligned}
Q\approx \varepsilon_q(\frac{3}{2}\partial_{v_y}^2
+\frac{3}{2}\partial_{v_z}^2-\partial_{v_x}^2)
\partial_{v_x}f(\textbf{r},\textbf{v},t)\,,
\end{aligned}
\end{equation}
with
\begin{equation}
\varepsilon_q=\frac{g\hbar^2}{4R^2m^2}=\frac{GM\hbar^2}{4R^4m^2}\,.
\end{equation}

Correspondingly, the function $f(\textbf{r},\textbf{v},t)$ can be regarded as a quasiprobability distribution in the phase space of position and velocity, with $P(\textbf{r},t)=\int_{-\infty}^{\infty}f(\textbf{r},\textbf{v},t){\rm d}^3v$ being a measurable probability distribution in real space.
With the given acceleration $\textbf{g}(\textbf{r})$, the distribution function $f(\textbf{r},\textbf{v},t)$ and its consequence $P(\textbf{r},t)$ depend on mass $m$ of test particles. This is obviously a nonclassical phenomenon. Note that, $Q$ decreases with increasing mass, the corresponding principle holds. This guarantees the validity of the above semiclassical approximation. One can find that the neglected terms in the original Wigner equation are relating to the high orders of small-quantity $\hbar^2/m^2$.
The nonzero $Q$ requires that third and higher derivatives of $V(\textbf{r})$ are nonzero.
Thus, for the linear or quadratic potential, the Wigner equation takes the same form as the classical Liouville equation. Nevertheless, we cannot conclude that the states within linear or quadratic potential are classical, because the initial states can be prepared in nonclassical states whose Wigner functions have negative values, for example, the well-known Fock state $|n\rangle$ (with $n\geq1$) of harmonic oscillator~\cite{WignerION,Squeeze}. Finally, we would like to emphasize that the Wigner equation (50) holds also for the two-body motion, where the variables $\textbf{r}$ and $\textbf{v}$ are respectively the relative position and relative velocity between two gravitationally interacting objects (of masses $m_1$ and $m_2$). The quantum correction is  $\varepsilon_q^{d}=\hbar^2 G M^3/(4R^4m_1^2m_2^2)$ with total mass $M=m_1+m_2$ and the reduced mass $m_1m_2/(m_1+m_2)$. The $\varepsilon_q^{d}$ can reduce to (52) with $m_1\gg m_2$.

\subsection{The perturbation solution}
In short, we rewrite the Wigner equation (50) as
\begin{equation}
\partial_t f=-(\hat{L}-\varepsilon_q\hat{L}_q)f\,,
\end{equation}
with the classical Liouville operator
\begin{equation}
\hat{L}=\textbf{v}\cdot\nabla_r+\textbf{g}(\textbf{r})\cdot
\nabla_v\,,
\end{equation}
and a quantum correction
\begin{equation}
\hat{L}_q=(\frac{3}{2}\partial_{v_y}^2
+\frac{3}{2}\partial_{v_z}^2-\partial_{v_x}^2)\partial_{v_x}\,.
\end{equation}

Similar to the solution of Schr\"{o}dinger equation in an interacting picture, we formally write the solution of Eq.~(53) as
\begin{equation}
f=e^{-t\hat{L}}f'\,,
\end{equation}
with
\begin{equation}
\begin{aligned}
\partial_t f'&=\varepsilon_qe^{t\hat{L}}\hat{L}_qe^{-t\hat{L}}f'
\\
&\approx\varepsilon_qe^{t\hat{L}_0}\hat{L}_qe^{-t\hat{L}_0}f'\,,
\end{aligned}
\end{equation}
and where
\begin{equation}
\hat{L}_0=v_x\partial_x+v_y\partial_y+v_z\partial_z-g\partial_{v_x}
\end{equation}
is just the Liouville operator with constant gravitational acceleration $g$.
For deriving the first line in (57), we have used the relation $\exp(t\hat{L})\exp(-t\hat{L})=1$, which can be easily proved by the Zassenhaus formula (24) in Sec. III. In the second line of (57), the gradient gravity in the exponential operator has been neglected, because $\varepsilon_q$ is already a small quantity. This approximation will greatly simplify our subsequent derivations.

Integrating Eq.~(57) and neglecting the high orders of $\varepsilon_q^2$, we have
\begin{equation}
\begin{aligned}
f'(\textbf{r},\textbf{v},t)&=f'(\textbf{r},\textbf{v},0)+\varepsilon_q\int_0^t
e^{\tau\hat{L}_0}\hat{L}_qe^{-\tau\hat{L}_0}f'(\textbf{r},\textbf{v},\tau){\rm d}\tau\\
&\approx f'(\textbf{r},\textbf{v},0)
+\varepsilon_q\int_0^te^{\tau\hat{L}_0}\hat{L}_qe^{-\tau\hat{L}_0}{\rm d}\tau\,\, f'(\textbf{r},\textbf{v},0)+\mathcal{O}(\varepsilon_q^2)\,.
\end{aligned}
\end{equation}
As a consequence, the Wigner function is solved as
\begin{equation}
\begin{aligned}
f(\textbf{r},\textbf{v},t)&\approx e^{-t\hat{L}}f'(\textbf{r},\textbf{v},0)+\varepsilon_q \hat{D}e^{-t\hat{L}_0}f'(\textbf{r},\textbf{v},0)\\
&=f_{c}(\textbf{r},\textbf{v},t)
+\varepsilon_q\hat{D}f_{u}(\textbf{r},\textbf{v},t)\\
&=f_{c}(\textbf{r},\textbf{v},t)+f_{q}(\textbf{r},\textbf{v},t)\,,
\end{aligned}
\end{equation}
with
\begin{equation}
\begin{aligned}
\hat{D}(t)
&=\int_0^{t}e^{(\tau-t)\hat{L}_0}\hat{L}_qe^{-(\tau-t)\hat{L}_0}{\rm d}\tau\\
&=
\int_{-t}^{0}e^{\tau\hat{L}_0}\hat{L}_qe^{-\tau\hat{L}_0}{\rm d}\tau\\
&=
\int_{-t}^{0}\hat{S}(\tau){\rm d}\tau.
\end{aligned}
\end{equation}
In Eq.~(60), $f_{c}(\textbf{r},\textbf{v},t)=\exp(-t\hat{L})f(\textbf{r},\textbf{v},0)$ and $f_{u}(\textbf{r},\textbf{v},t)=\exp(-t\hat{L}_0)f(\textbf{r},\textbf{v},0)$ are the solutions of classical Liouville equations with nonuniform and the uniform gravitational accelerations, respectively. The quantum mechanical correction $f_q(\textbf{r},\textbf{v},t)=\varepsilon_q \hat{D}f_{u}(\textbf{r},\textbf{v},t)$ is equal to zero at the initial time $t=0$, so that $f_q(\textbf{r},\textbf{v},t)$ is not the correction for the initial state but a dynamical one.

The integral kernel $\hat{S}(\tau)=\exp(\tau\hat{L}_0)\hat{L}_q\exp(-\tau\hat{L}_0)$ in (61) can be computed by using BCH formula (25) in Sec. III. We have
\begin{equation}
\begin{aligned}
e^{\tau\hat{L}_0}\partial_{v_x}e^{-\tau\hat{L}_0}
=\partial_{v_x}+\tau[\hat{L}_0,\partial_{v_x}]=\partial_{v_x}-\tau\partial_x\,,
\end{aligned}
\end{equation}
and consequently
\begin{equation}
\begin{aligned}
e^{\tau\hat{L}_0}\partial_{v_x}^2e^{-\tau\hat{L}_0}
&=e^{\tau\hat{L}_0}\partial_{v_x}
e^{-\tau\hat{L}_0}e^{\tau\hat{L}_0}\partial_{v_x}e^{-\tau\hat{L}_0}\\
&=(\partial_{v_x}-\tau\partial_x)^2.
\end{aligned}
\end{equation}
These formulas can be directly generalized into $y$- and $z$-directions, and thus
\begin{equation}
\begin{aligned}
\hat{S}(\tau)
&=\frac{3}{2}
(\partial_{v_y}-\tau\partial_y)^2(\partial_{v_x}-\tau\partial_x)\\
&\,\,\,
+\frac{3}{2}
(\partial_{v_z}-\tau\partial_z)^2(\partial_{v_x}-\tau\partial_x)\\
&\,\,\,
-(\partial_{v_x}-\tau\partial_x)^3\,.
\end{aligned}
\end{equation}

Operator (64) includes many terms, so that solving the quantum correction $f_q(\textbf{r},\textbf{v},t)
=\varepsilon_q\int_{-t}^0\hat{S}(\tau)f_u(\textbf{r},\textbf{v},t){\rm d\tau}$ is still very complex. However, it is easy to solve the quantum correction in one-dimensional real space, for example,
\begin{equation}
\begin{aligned}
P_q(x,t)&=\int_{-\infty}^\infty f_{q}(\textbf{r},\textbf{v},t){\rm d}^3 v{\rm d}y{\rm d}z\\
&=\varepsilon_q\int_{-t}^0\int_{-\infty}^\infty \hat{S}(\tau)
f_{u}(\textbf{r},\textbf{v},t){\rm d}^3 v{\rm d}y{\rm d}z{\rm d}\tau\\
&=-\frac{\varepsilon_qt^4}{4}\partial^3_x\int_{-\infty}^\infty
f_{u}(\textbf{r},\textbf{v},t){\rm d}^3 v{\rm d}y{\rm d}z\\
&=-\frac{\varepsilon_qt^4}{4}\partial^3_x P_u(x,t)\,.
\end{aligned}
\end{equation}
In the third line of the above equation, we have used the locality of classical particles ensemble,
\begin{equation}
\begin{aligned}
{\lim_{\eta \to \pm\infty}}f_u(\textbf{r},\textbf{v},t)={\lim_{\eta \to \pm\infty}}[\partial_\eta f_u(\textbf{r},\textbf{v},t)]={\lim_{\eta \to \pm\infty}}[\partial_\eta^2f_u(\textbf{r},\textbf{v},t)]=0\,,
\end{aligned}
\end{equation}
with $\eta$ being one of the variables $(x,y,z,v_x,v_y,v_z)$, and thus $\int_{-\infty}^{\infty}(\partial_\eta f_u){\rm d}\eta=\int_{-\infty}^{\infty}(\partial_\eta^2f_u){\rm d}\eta=\int_{-\infty}^{\infty}(\partial_\eta^3f_u){\rm d}\eta=0$.
In the last line of Eq.~(65), $P_u(x,t)=\int_{-\infty}^\infty
f_{u}(\textbf{r},\textbf{v},t){\rm d}^3 v{\rm d}y{\rm d}z$ is the one-dimensional probability distribution with constant acceleration.
For the same reason, $\int_{-\infty}^\infty P_q(x,t){\rm d}x=0$, so that the total probability is conserved (the classical distribution $f_c$ in Eq. (60) is already normalized).

Note that, Eq.~(65) can be proved by directly applying the Weyl transform to the already obtained Heisenberg operator $\hat{x}(t)$ in Sec.~II. The Weyl transforms of $\hat{x}(t)$ and $\hat{x}(t)^2$ are their original functions, so that $\langle i|\hat{x}(t)|i\rangle$ and $\langle i|\hat{x}(t)^2|i\rangle$ evolve classically. This means that the above $P_q(x,t)$ does not contribute to the average values of position and its square, i.e., $\int_{-\infty}^{\infty}x\partial_x^3P_u{\rm d}x=\int_{-\infty}^{\infty}x^2\partial_x^3P_u{\rm d}x=0$. This can be proved by expanding the integrands as the forms of first-order derivation: $x\partial_x^3P_u=\partial_x(x\partial_x^2P_u-\partial_xP_u)$, $x^2\partial_x^3P_u=\partial_x(x^2\partial_x^2P_u-2x\partial_xP_u
+2P_u)$. The functions in the round brackets are convergent at $x\rightarrow \pm\infty$ for any local distribution $P_u$. In fact, the above zero result can be also proved by using the Dirac delta function. Writing
$P_u(x,t)=\int_{-\infty}^{\infty}P_u(x',t)\delta(x-x'){\rm d}x'$, we have $\int_{-\infty}^{\infty}x^n\partial_x^3P_u(x,t){\rm d}x=\int_{-\infty}^{\infty}\int_{-\infty}^{\infty}P_u(x',t)x^n\partial_x^3
\delta(x-x'){\rm d}x{\rm d}x'=-\int_{-\infty}^{\infty}P_u(x',t)(\partial_{x'}^3x'^{n}){\rm d}x'$, with $n=0,1,2,3$. Immediately, we see $\int_{-\infty}^{\infty}x^3\partial_x^3P_u{\rm d}x=-6$, i.e., the quantum correction should contribute the average value of $x^3$. This is true, and can be also proved by the Weyl transform of cubic operator $\hat{x}(t)^3$. Using formula (19) in Sec. II, one can find $\hat{x}(t)^3$ containing $\hat{x}^2\hat{v}_x^2+\hat{v}_x^2\hat{x}^2$, $\hat{x}\hat{v}_x^2\hat{x}$, $\hat{v}_x\hat{x}^2\hat{v}_x$, and $(\hat{x}\hat{v}_x+\hat{v}_x\hat{x})^2$, whose Weyl transforms are not their original functions~\cite{AmJP}.

\subsection{The tiny quantum fluctuation}
The quantum correction $P_q(x,t)$ is now resolvable based on the given classical probability density $P_u(x,t)=\int_{-\infty}^\infty
f_{u}(\textbf{r},\textbf{v},t){\rm d}^3 v{\rm d}y{\rm d}z$ in the uniform gravitational field, where the integrand can be easily computed by the classical trajectory-dynamics method~\cite{Delta},
\begin{equation}
\begin{aligned}
f_{u}(x,v_x,t)&=\int_{-\infty}^{\infty}
\delta\left[x-(x_i+v_{xi}t-gt^2/2),v_x-(v_{xi}-gt)\right]f(x_i,v_{xi},0){\rm d}x_i{\rm d}v_{xi}\\
&=f_u(x-v_xt-gt^2/2,v_x+gt)\,.
\end{aligned}
\end{equation}
In short, the variables in $y$- and $z$-directions are not written, and $f(x_i,v_{xi},0)$ is the probability of particles that are initially at position $x_i$ and velocity $v_{xi}$.
Supposing the initial Wigner function takes Gaussian form~\cite{PRDG}:
\begin{equation}
\begin{aligned}
f(x,v_x,0)=\frac{1}{2\pi\sigma_x\sigma_v}
\exp\left[-\frac{x^2}{2\sigma_x^2}-\frac{v_x^2}{2\sigma_v^2}\right]\,,
\end{aligned}
\end{equation}
with the standard deviations $\sigma_x$ and $\sigma_v$. According to Eq.~(67), replacing $x$ by $x-v_xt-gt^2/2$, and $v_x$ by $v_x+gt$, the above initial state becomes the desired time-dependent state $f_{u}(x,v_x,t)$. Consequently, using Gaussian integral we have
\begin{equation}
\begin{aligned}
P_{u}(x,t)&=\int_{-\infty}^{\infty}f_{u}(x,v_x,t) {\rm d}v_x\\
&=\frac{1}{\sqrt{2\pi}\sqrt{\sigma_x^2+\sigma_v^2t^2}}
\exp\left[-\frac{1}{2}\xi^2(x,t)\right]\,,
\end{aligned}
\end{equation}
with
\begin{equation}
\begin{aligned}
\xi(x,t)=\frac{x+\frac{1}{2}gt^2}{\sqrt{\sigma_x^2+\sigma_v^2t^2}}\,.
\end{aligned}
\end{equation}
Finally, quantum probability density (65) is given as:
\begin{equation}
\begin{aligned}
P_q(x,t)=P_0\frac{t^4\left[\xi^3(x,t)-3\xi(x,t)\right]}
{\left[(\sigma_x/\sigma_v)^2+t^2\right]^2\sqrt{2\pi}}
\exp\left[-\frac{1}{2}\xi^2(x,t)\right]\,,
\end{aligned}
\end{equation}
with
\begin{equation}
\begin{aligned}
P_0=\frac{g\hbar^2}{16R^2m^2\sigma_v^4}\,.
\end{aligned}
\end{equation}
\begin{figure}[tbp]
\includegraphics[width=9cm]{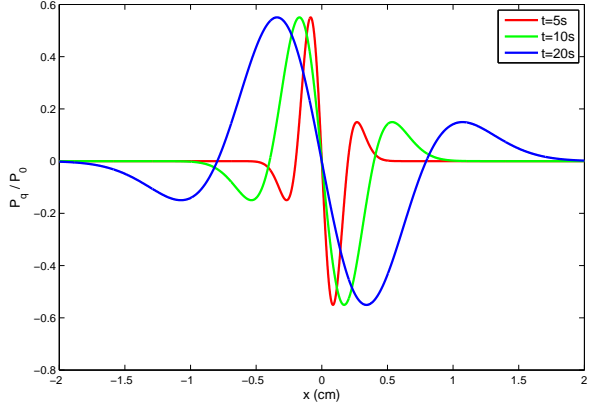}
\caption{Quantum fluctuation in one-dimensional real space, with $R=1.5\,\text{m}$, $M=10^3\,\text{kg}$, $m=10^{-25}$~kg, $\sigma_x\approx3\,\mu\text{m}$, and $\sigma_v\approx3\times10^{-4}\,\text{m}/\text{s}$.}
\end{figure}

Fig.~3 shows that $P_q(x,t)$ is positive in some regions and negative in others. The previous arguments $\int_{-\infty}^{\infty} P_q(x,t){\rm d}x=0$ etc. can be also proved by the numerical integrals of (71).
The $P_0$ is the characteristic probability density, which increases rapidly with the decreasing of velocity uncertainty, as $1/\sigma_v^4$. Thus, the cooled neutral atom may be appropriate for achieving a relatively large value of $P_0$. Consider an atom (of mass $m$) is initially confined in optical trap as a harmonic oscillator (of frequency $\omega$) and laser cooled in its vibrational ground state. Removing instantaneously the optical trapping, atom is released as a freely falling body. The position and velocity uncertainties of the initial wave packet are on the orders of $\sigma_x=\sqrt{\hbar/(2m\omega)}\approx3\,\mu\text{m}$ and $\sigma_v=\sqrt{\hbar \omega/(2m)}\approx0.3\,\text{mm}/\text{s}$, if $m\approx10^{-25}$~kg and $\omega\approx100$~Hz~\cite{cooling1,cooling2}. For the term $g/R^2=GM/R^4$, we still consider $R=1.5\,\text{m}$ and $M=10^3\,\text{kg}$ as before, and thus $P_0\approx3.3\times10^{-13}$ per meter. This is a very small probability density, beyond the experimental ability. Assuming  $\sigma_v\approx2.3\,\mu\text{m}/\text{s}$ and $\sigma_x\approx0.23\,\text{mm}$ of $\omega=0.01~\text{Hz}$ (not the experimental data), then $P_0$ is significantly enlarged to $P_0\approx3.3\times10^{-5}$ per meter.

One may notice $P_0\propto1/m^2$ and thus electrons should have significant quantum effects. This is true and can be found in the hydrogen atom. If we replace electronic mass by an atomic mass and retain Coulomb potential unchanging, then the separateness of the Bohr radius is not obvious. However, using electron as a free fall to measure gravity is impractical, because it is very sensitive to the electromagnetic noise. Thus, the key technologies for quantum free fall experiments are atom cooling and noises shielding.
Finally, we emphasize that $P_q$ is still not the final result. The true probability density is $P=P_c+P_q$, see Eq.~(60). The classical part $P_c$ may be numerically solved by the computer command of conditional sum: $P_c(x,t)=\prod_j
\sum_{n_j}f(\textbf{r}_d,\textbf{v}_d,0)(\Delta x)^2
(\Delta v)^3
\,,\,\,{\rm if}\,\,\,\,x\leq X(\textbf{r}_d,\textbf{v}_d,t)\leq x+\Delta x$.
Here, $f(\textbf{r}_d,\textbf{v}_d,0)$ is the initial distribution function with the discretized position $\textbf{r}_d=(n_1,n_2,n_3)\Delta x$ and velocity $\textbf{v}_d=(n_4,n_5,n_6)\Delta v$, and $X(\textbf{r}_d,\textbf{v}_d,t)$ is the numerical solution of the Newtonian equation (15), with any initial position $\textbf{r}_d$ and initial velocity $\textbf{v}_d$. As mentioned before, if one neglects the present term $P_q$, there still exists quantum phenomena in $P_c$ (such as the familiar  double-slit interference of matter waves in flat spacetime), because the initial state $f(\textbf{r}_d,\textbf{v}_d,0)$ can be nonclassical. Hence, a more precise statement for $P_c(x,t)$ is that, it can be solved by the classical trajectory with any given initial Wigner function.

\section{Conclusion}
We have studied the free fall of microparticles with high-order gravity gradients. It is shown that, the cubic terms in Newtonian potential shall generate a new phase shift in AI, which depends on the position and velocity uncertainties of the incident atoms. Certainly, this effect is negligible in the ground-based laboratory, because Earth's radius is far larger than atomic freely falling length limited by the practical vacuum installation. However, the present effect may be considerable in the space laboratory due to the gravity of the satellite. On the one hand, the size of satellite is far smaller than Earth. On the other hand, the freely falling time can be very long within the microgravity system. Thus, this study may be useful for designing the high-precision AIs of avoidable systematic effects of nonlinear gravity. For the sake of economy, we also suggested using the force of atom moving in the nonuniform magnetic field to simulate the nonlinear gravity in the ground-based laboratory.

Another value of the present study, which is purely theoretical, refers to the  nonclassicality of freely falling particles. With the presence of high-order gravity gradients, there exists a quantum correction in the dynamical equation of the Wigner function. Certainly, its contribution to the observable value is very tiny and has less feasibility in the current experiments.
Nevertheless, we have learned that the nonclassicality of free fall refers two aspects. One is the initial state which can be nonclassical, prepared by the nongravity interaction. The other is the quantum-corrected dynamical equation regardless of initial state. The two aspects are both mass dependent.
Thus, the freely falling microparticles should be mass dependent in general. As mentioned earlier, this is just a quantum mechanical effect, not EP violation. On the one hand, the dynamical equations throughout the paper are built on the principle that inertia mass is equal to gravitational mass. On the other hand, the gravitational Wigner equation satisfies the corresponding principle that the quantum correction vanishes with the increasing mass of test particles. Hence, the present violation of the universality of free fall is just a quantum mechanical effect of microparticles, not macroparticles.

\textbf{Acknowledgments}:
This work was supported partly by the National Natural Science Foundation of China, Grants No. 11747311, No. 61871333, and No. 11204249.

\end{document}